\documentstyle[aps,epsf,graphicx]{revtex}\tighten

\begin{document}

\title{Gravitational waves in preheating}

\author{Daniel Tilley and Roy Maartens}

\address{Relativity and Cosmology Group, Division of Mathematics and
Statistics, Portsmouth University, Portsmouth~PO1~2EG, Britain}

\maketitle

\begin{abstract}
We study the evolution of gravitational waves through the
preheating era that follows inflation. The oscillating inflaton
drives parametric resonant growth of scalar field fluctuations,
and although super-Hubble tensor modes are not strongly amplified,
they do carry an imprint of preheating. This is clearly seen in
the Weyl tensor, which provides a covariant description of
gravitational waves.

\end{abstract}

\pacs{9880 \hfill gr-qc/0002089}

\section*{1. Introduction}

Gravitational waves in inflationary cosmology are produced by
sub-Hubble scale vacuum fluctuations. They are stretched to
super-Hubble scales by inflationary expansion, and then they
generate anisotropies in the cosmic microwave background (CMB)
(see, e.g., Grishchuk 1975, Mukhanov et al. 1992). Scales which
re-enter before matter-radiation equality will rapidly redshift
(and suffer small damping during recombination), having negligible
effect on CMB temperature anisotropies, while scales which
re-enter later can affect the temperature anisotropies at large
angles. Roughly speaking, we can approximate the impact of
gravitational waves upon CMB temperature anisotropies by
neglecting all local causal dynamics and treating the large-scale
fluctuations as frozen beyond the Hubble scale until re-entry,
where their amplitude is conserved and determined by the
inflationary Hubble rate at the time of Hubble-crossing. This
picture is confirmed by the evolution equation (Lifshitz 1946)
\begin{equation}
f''_{ij}+2{a'\over a}f'_{ij}+k^2f_{ij}=0
 \label{1}
\end{equation}
for the Fourier modes of the transverse traceless tensor
perturbation $f_{ij}$, defined (on a flat background) by
\begin{equation}
ds^2=a^2(\eta)\left[-d\eta^2+\left(\delta_{ij}+
f_{ij}\right)dx^idx^j\right]\,.
 \label{2}
\end{equation}
Equation (\ref{1}) shows that on very large scales, $k/aH\ll 1$
(where $H=a'/a^2$),
\begin{equation}
f_{ij}\approx A_{ij}+B_{ij}\int {d\eta\over a^2}\,,
 \label{3}
\end{equation}
where $A'_{ij}=0=B'_{ij}$. If $a'>0$ then the $B$-mode is
decaying, and we have $f_{ij}\approx$ constant, with the constant
determined by the Hubble rate at inflationary Hubble-crossing,
$H(\eta_{c})=k/a(\eta_{c})$.

This simple picture is modified by small corrections induced at
preheating. At the end of slow-roll inflation, the inflaton
oscillates and transfers its energy to fluctuations, initiating
the reheating era, which ends when created particles thermalize as
a radiative plasma. Reheating is often initiated by a preheating
era, marked by coherent inflaton oscillations which drive
parametric resonant amplification of scalar fluctuations (see,
e.g., Traschen and Brandenberger 1990, Kofman et al. 1994, 1997).
One crucial point about preheating is that, since the inflaton is
coherent on scales well beyond the Hubble horizon, it is in
principle possible for super-Hubble fluctuations to be amplified
{\em without violating causality} (Bassett et al.
1999a,b).\footnote{It is also possible for small-scale
gravitational waves to be generated by gravitational bremstrahlung
via rescattering of scalar fluctuations during preheating
(Khlebnikov and Tkachev 1997). This is a particular example of the
generation of tensor perturbations by scalar perturbations at
second order (Matarrese et al. 1998). } In the case of scalar
fluctuations of the {\em metric}, this can in principle produce
nonlinear amplification, depending on initial conditions and
coupling strengths (Bassett et al. 1999a,b,c, 2000, Ivanov 2000,
Jedamzik and Sigl 2000, Liddle et al. 2000). While tensor
fluctuations will not be strongly amplified by scalar inflaton
oscillations, these oscillations nevertheless could leave an
imprint on the tensor spectrum on large scales. Such a possibility
is usually ruled out on causality grounds, but no such causality
constraint operates during coherent oscillations of the inflaton
condensate.

These small corrections will be carried into Eq. (\ref{1}) via the
scale factor, which inherits an oscillatory addition to its
average value. However, the nature of the effect is more clearly
brought out via an alternative description of gravitational waves,
based on the idea that a full characterization requires the
curvature tensor, not the metric (Pirani 1957). Transverse
traceless modes are given by the divergence-free electric and
magnetic parts of the Weyl tensor
\[
E_{\mu\nu}=C_{\mu\alpha\nu\beta}u^\alpha u^\beta\,,~~
H_{\mu\nu}=^*\!\!C_{\mu\alpha\nu\beta}u^\alpha u^\beta\,,
\]
where $u^\mu$ is the background four-velocity (there are no
velocity perturbations for tensor modes). This covariant
description of gravitational waves was used by Hawking (1966), and
is remarkably analogous to electromagnetic radiation theory
(Dunsby et al. 1997, Maartens and Bassett 1998).

In this paper, we use the Maxwell-Weyl approach to gravitational
waves to investigate the effects of inflaton oscillations in some
simple preheating models, generalizing previous work in Minkowski
spacetime (Bassett 1997). In Sections 2 and 3, we give the basic
equations and their qualitative properties. In Section 4 we
present the numerical calculations, and Section 5 contains
concluding remarks and discussion.

\section*{2. Background dynamical equations}

The background inflaton is governed by the Klein-Gordon equation
\begin{equation}\label{4}
\ddot{\varphi}+3H\dot{\varphi}+V_\varphi=0\,,
\end{equation}
where $V_\varphi=\partial V/\partial\varphi$. We will consider the
simple chaotic inflation potentials $V={1\over2}m^2\varphi^2$ and
$V={1\over4}\lambda\varphi^4$. Although the resonance in scalar
fluctuations is dramatically increased when the inflaton is
coupled to other fields, for example via the additional potential
term ${1\over2}g^2\varphi^2\chi^2$, this does not seem to have an
effect on {\em tensor} fluctuations (Tilley 2000). Thus we will
confine ourselves to simplified single-field models of preheating.
Preheating in more realistic models ends when backreaction effects
of the fluctuations destroy the coherence of inflaton
oscillations. In our simplified models, backreaction effects are
not incorporated, but we can use the results from detailed
investigations to estimate the time that preheating lasts (see,
e.g., Kofman et al. 1997).

The Hubble rate is determined by the Friedmann equation
\begin{equation}\label{5}
H^2={\textstyle{1\over3}}\kappa^2\left[{\textstyle{1\over2}}
\dot{\varphi}^2+V(\varphi)\right]\,,
\end{equation}
where $\kappa^2=8\pi/M_{p}^2$. Equations (\ref{4}) and (\ref{5})
imply $\dot{H}=-{1\over2}\kappa^2\dot{\varphi}^2$. The energy
density and effective pressure of the inflaton are
\begin{equation}\label{6}
\rho=\kappa^2\left({\textstyle{1\over2}}\dot{\varphi}^2+V\right)\,,~~
p=\kappa^2\left({\textstyle{1\over2}}\dot{\varphi}^2-V\right)\,.
\end{equation}

During slow-roll inflation, the coupled equations (\ref{4}) and
(\ref{5}) have approximate analytic solutions for the simple
potentials. When the value of $\varphi$ drops low enough,
slow-roll ends and the oscillatory regime begins. The approximate
analytic forms are (see, e.g., Kaiser 1997, Kofman et al. 1997)
\begin{eqnarray}
\varphi &\approx& \varphi_{in}{\sin\tau\over\tau} ~~\mbox{where}~~
\tau=m(t-t_{in}) ~\mbox{and}~
V={\textstyle{1\over2}}m^2\varphi^2\,,\label{7}\\ \varphi &
\approx &{\varphi_{in}\over a} {\mbox{cn}}\,\left( \tau,
{1\over\sqrt{2}}\right)~~\mbox{where}~~
\tau=\sqrt{\lambda}\varphi_{in}(\eta-\eta_{in})~\mbox{and}~
V={\textstyle{1\over4}}\lambda\varphi^4\,,\label{8}
\end{eqnarray}
where $a_{in}=1$ and cn is a Jacobian elliptic function. The
initial values of $\varphi_{in}$ are $\sim 0.3M_{p}$ in the
quadratic case, and $\sim 0.6M_{p}$ in the quartic case.

Time-averaging over oscillations shows that $\bar{a}\propto
t^{2/3}$ for the quadratic potential, and $\bar{a}\propto t^{1/2}$
for the quartic potential. These are the asymptotic values of the
scale factor, but in practice backreaction effects due to
couplings will end the preheating oscillations. If one uses the
average forms $\bar{a}$ for $a$, i.e., if one ignores oscillatory
behaviour in the inflaton, then one regains the standard results
for gravitational wave evolution in the dust and radiation eras.
In particular, models with a smooth transition from inflation to
radiation-domination, neglecting reheating dynamics, show that
there is no super-Hubble amplification of gravitational waves
(e.g. Caldwell 1996, Tilley and Maartens 1998) . Our numerical
integrations show that this averaging loses interesting features
in the gravitational waves (see Section 3). We do not use the
approximate forms in Eqs. (\ref{7}) and (\ref{8}) for our
numerical results---instead, we integrate the Friedmann and
Klein-Gordon equations numerically.

\section*{3. Maxwell-Weyl gravitational wave equations}

Gravitational wave perturbations in the covariant approach are
governed by the Maxwell-Weyl equations:
\begin{eqnarray*}
&&\left(\mbox{div}\,E\right)_\mu=0=\left(\mbox{div}\,H\right)_\mu
\,,\\ &&\dot{E}_{\mu\nu}+3HE_{\mu\nu}-\mbox{curl}\,H_{\mu\nu}=-
{\textstyle{1\over2}}(\rho+p)\sigma_{\mu\nu}\,,\\
&&\dot{H}_{\mu\nu}+3HH_{\mu\nu}+\mbox{curl}\,E_{\mu\nu} =0\,,
\end{eqnarray*}
where $\sigma_{\mu\nu}$ is the shear, a dot denotes
$u^\mu\nabla_\mu$, and div and curl are the covariant spatial
divergence and curl for tensors (Maartens and Bassett 1998). These
equations hold for perfect fluids and minimally coupled scalar
fields (in which case $\rho+p=\kappa^2\dot{\varphi}^2$). The shear
is a tensor potential for the electric and magnetic Weyl tensors
(Maartens and Bassett 1998):
\begin{eqnarray}
E_{\mu\nu} &=&
-\dot{\sigma}_{\mu\nu}-2H\sigma_{\mu\nu}\,,\label{9}\\ H_{\mu\nu}
&=& \mbox{curl}\,\sigma_{\mu\nu} \,,\label{10}
\end{eqnarray}
in close analogy with the Maxwell relations
$\vec{E}=-\dot{\vec{A}}$ and $\vec{H}=\mbox{curl}\,\vec{A}$.
Taking the curl and dot of the Maxwell-Weyl propagation equations,
and using the identity for the tensor curl of the curl, we find
wave equations for the three tensors (Dunsby et al. 1997, Maartens
and Bassett 1998, Challinor 2000). Using equations (\ref{4}) and
(\ref{5}), these become
\begin{eqnarray}
\ddot{E}_{\mu\nu}+7H\dot{E}_{\mu\nu}+4\kappa^2V(\varphi)E_{\mu\nu}
&=& \Delta E_{\mu\nu}+\kappa^2\dot{\varphi}\left[5H\dot{\varphi}
+V'(\varphi)\right]\sigma_{\mu\nu} \,,\label{11}\\
\ddot{H}_{\mu\nu}+7H\dot{H}_{\mu\nu}+4\kappa^2V(\varphi)H_{\mu\nu}
&=& \Delta H_{\mu\nu}\,,\label{12}
\end{eqnarray}
where $\Delta$ is the covariant spatial Laplacian. One can also
derive a wave equation for the shear:
\begin{equation}\label{13}
\ddot{\sigma}_{\mu\nu}+5H\dot{\sigma}_{\mu\nu}+\kappa^2\left[2V(\varphi)-
{\textstyle{1\over2}}\dot{\varphi}^2\right]\sigma_{\mu\nu} =\Delta
\sigma_{\mu\nu}\,.
\end{equation}

We decompose the tensors into modes (Challinor 2000)
\begin{eqnarray}
E_{\mu\nu} &=& a^{-2}\sum k^2\left[{\cal E}_kQ_{\mu\nu}^{(k)}+
\tilde{\cal E}_k\tilde{Q}_{\mu\nu}^{(k)}\right]\,,\label{14}\\
H_{\mu\nu} &=& a^{-2}\sum k^2\left[{\cal H}_kQ_{\mu\nu}^{(k)}+
\tilde{\cal H}_k\tilde{Q}_{\mu\nu}^{(k)}\right]\,,\label{15}\\
\sigma_{\mu\nu} &=& a^{-1}\sum k\left[{\cal S}_kQ_{\mu\nu}^{(k)}+
\tilde{\cal S}_k\tilde{Q}_{\mu\nu}^{(k)}\right]\,,\label{16}
\end{eqnarray}
where $\sum$ denotes a symbolic sum over harmonic modes, and $Q$,
$\tilde{Q}$ are tensor harmonics of electric and magnetic parity,
which are time-independent, transverse traceless, and related by
\[
\mbox{curl}\,Q_{\mu\nu}^{(k)}={k\over
a}\,\tilde{Q}_{\mu\nu}^{(k)}\,,
\]
showing that the different polarization states are coupled. The
mode functions ${\cal E}_k$ determine the tensor contribution to
the CMB power spectrum. The magnetic Weyl mode functions are
algebraically related to the shear mode functions by
\begin{equation}\label{17}
{\cal H}_k=\tilde{\cal S}_k\,.
\end{equation}
Specializing the results in Challinor (2000) to the scalar field
case, we find the evolution equations
\begin{eqnarray}
\dot{\cal E}_k+3H{\cal E}_k-{a\over k}\left({k^2\over a^2}-
{\textstyle{1\over2}}\kappa^2\dot{\varphi}^2\right){\cal S}_k
&=&0\,,\label{18}\\ \dot{\cal S}_k+H{\cal S}_k+{k\over a}{\cal
E}_k &=& 0\,. \label{19}
\end{eqnarray}

For comparison, in the time-averaged approximation, the
$\dot{\varphi}^2$ term in the coefficient of ${\cal S}_k$ in Eq.
(\ref{18}) is replaced by $\gamma\rho$, where $\rho=\rho_{\rm
in}a^{-3\gamma}$, with $\gamma=1$ for the averaged quadratic case
(i.e. cold matter or `dust'), and $\gamma={4\over3}$ for the
averaged quartic case (i.e. radiation).

\section*{4. Numerical results}

Equations (\ref{17})--(\ref{19}) for the gravitational wave
perturbations, and (\ref{4})--(\ref{5}) for the background are
integrated numerically, and the results are compared with those
for the time-averaged approximation. We investigated the evolution
of COBE modes, which left the Hubble radius at about 50 to 60
e-folds before the end of inflation, so that $k/aH\sim 10^{-24}$
at the start of preheating, $t=t_{in}$. We also integrated for a
typical small scale mode, with $k/aH \sim 10$ at $t=t_{in}$. The
initial values for $\varphi$ were given in the previous section.
We used the slow-roll relation to set the initial inflaton
velocity, $\dot{\varphi}_{in}=-M_{p}V_\varphi/\sqrt{24\pi V}$. For
the tensor modes, we used the initial value $10^{-5}$. The
inflaton mass and self-coupling were chosen as $m=10^{-6}M_{p}$
and $\lambda=10^{-12}$. In order to take account of backreaction
effects, which will end the coherent inflaton oscillations, we
terminated the integrations after $\tau=100$.

The results are shown in Figs. 1--8.

\begin{figure}
\includegraphics[height=3.2in,width=3.2in]{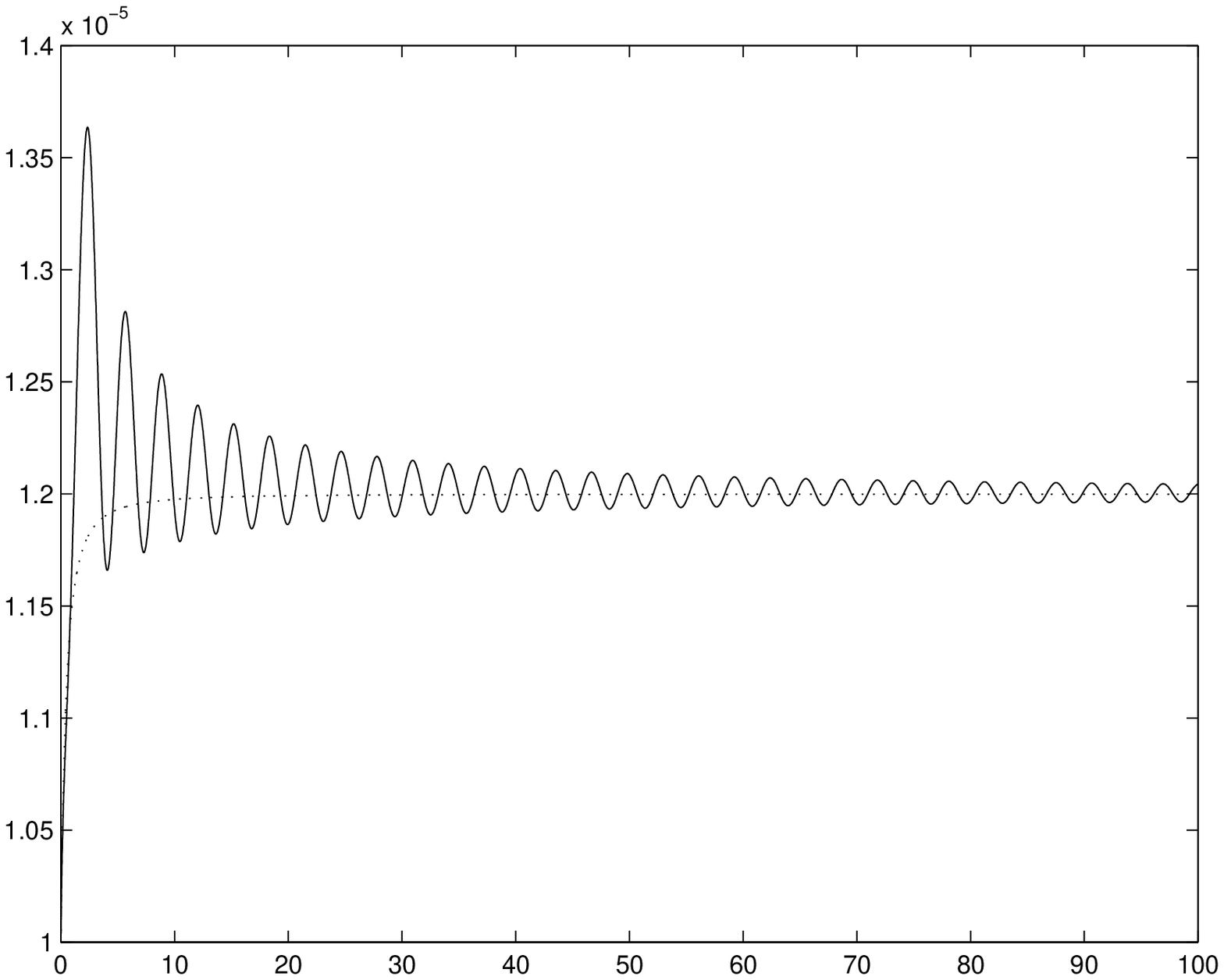}
\caption{The electric Weyl mode ${\cal E}_k(\tau)$ on super-Hubble
scales, $k/(aH)_{in}= 10^{-24}$. The solid line is for the
inflaton potential $V={1\over2}m^2\varphi^2$, while the dotted
line is for the corresponding time-averaged model, which behaves
like cold matter (dust).}
\end{figure}
\begin{figure}
\includegraphics[height=3.2in,width=3.2in]{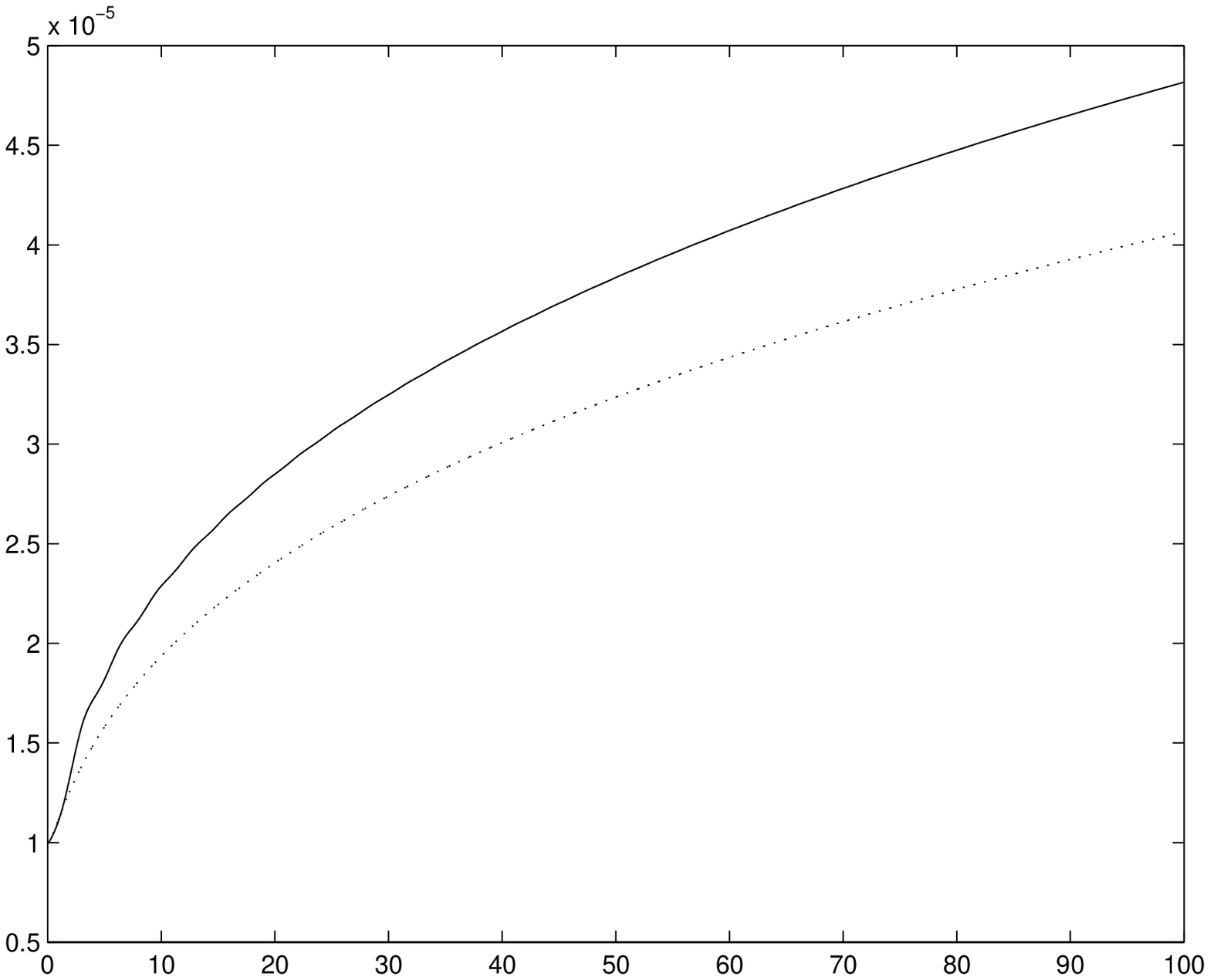}
\caption{The magnetic Weyl mode ${\cal H}_k(\tau)$, as for Fig.
1.}
\end{figure}
\begin{figure}
\includegraphics[height=3.2in,width=3.2in]{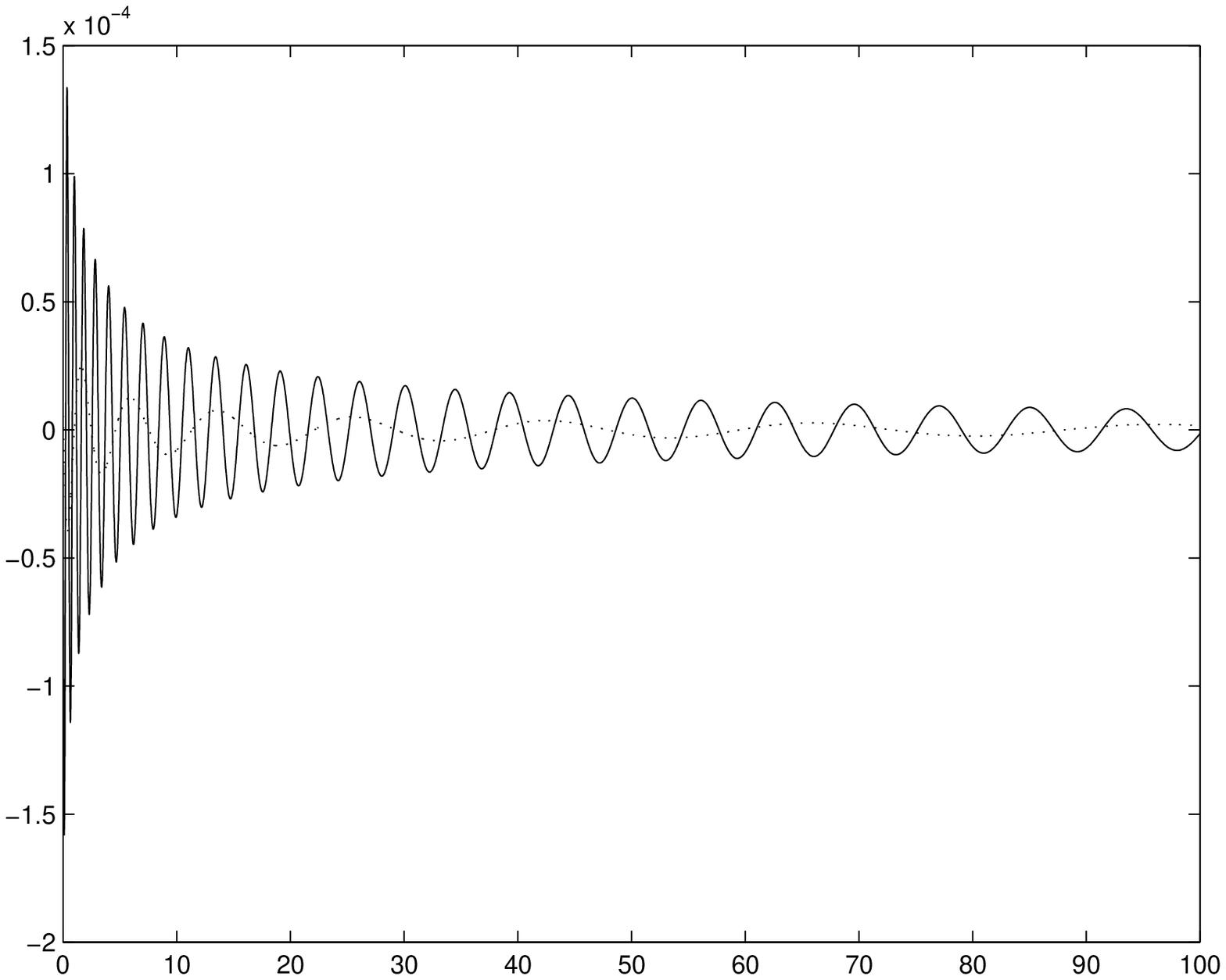}
\caption{The electric Weyl mode ${\cal E}_k(\tau)$ on sub-Hubble
scales, $k/(aH)_{in}= 10$. The solid line is for the inflaton
potential $V={1\over2}m^2\varphi^2$, while the dotted line is for
the corresponding time-averaged model, which behaves like cold
matter (dust).}
\end{figure}
\begin{figure}
\includegraphics[height=3.2in,width=3.2in]{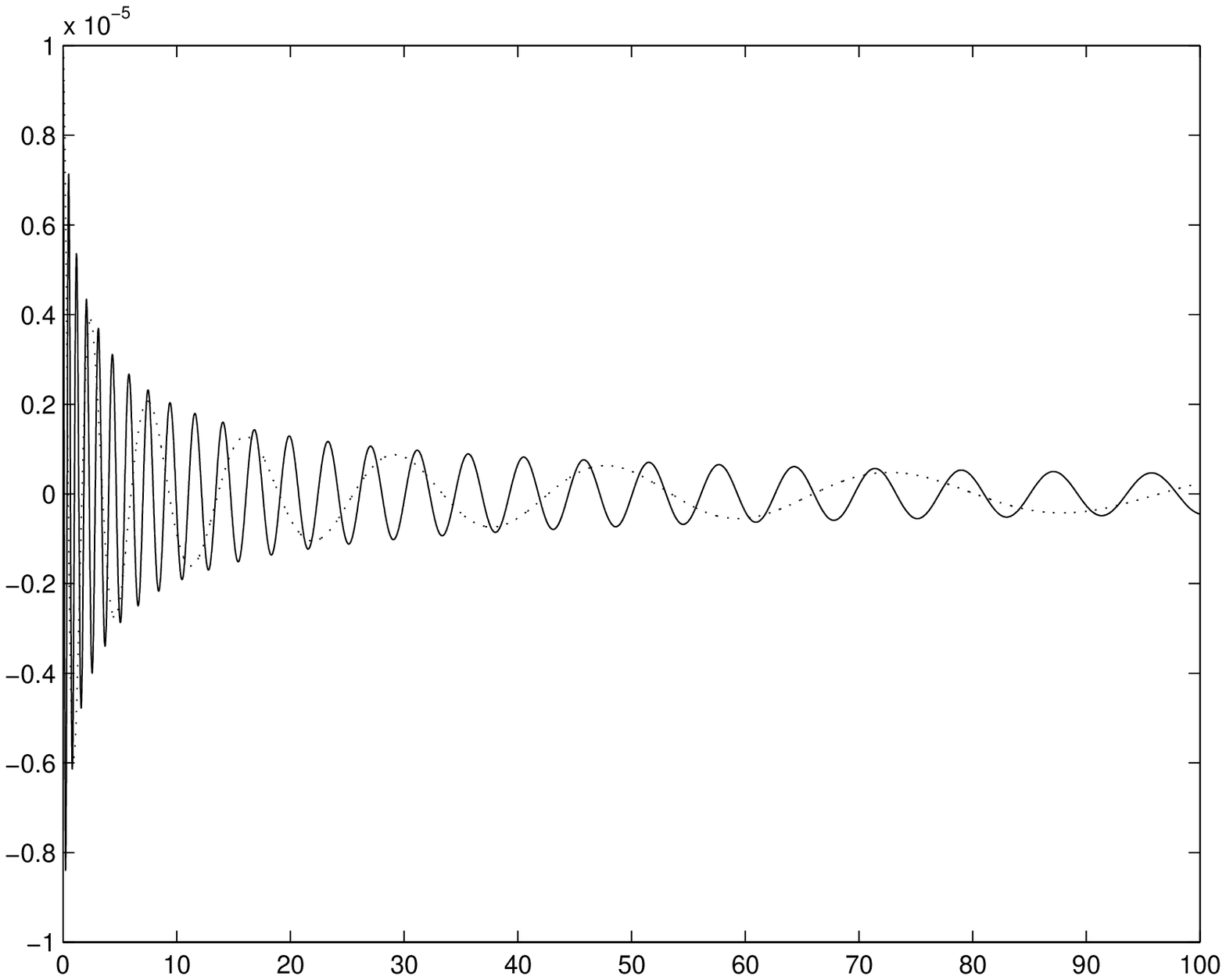}
\caption{The magnetic Weyl mode ${\cal H}_k(\tau)$, as for Fig.
3.}
\end{figure}
\begin{figure}
\includegraphics[height=3.2in,width=3.2in]{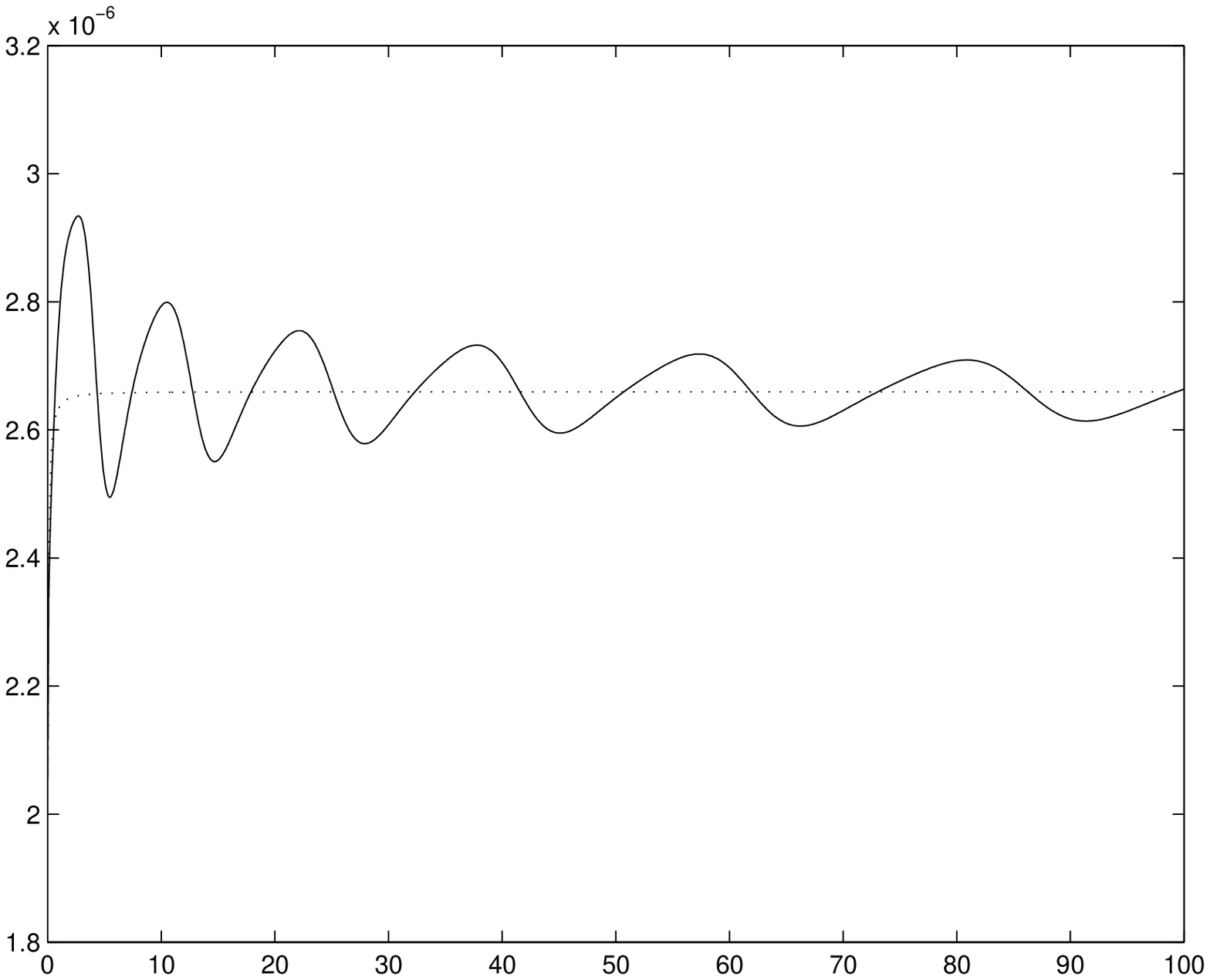}
\caption{The electric Weyl mode ${\cal E}_k(\tau)$ on super-Hubble
scales, $k/(aH)_{in}= 10^{-24}$. The solid line is for the
inflaton potential $V={1\over4}\lambda^4\varphi^4$, while the
dotted line is for the corresponding time-averaged model, which
behaves like radiation.}
\end{figure}
\begin{figure}
\includegraphics[height=3.2in,width=3.2in]{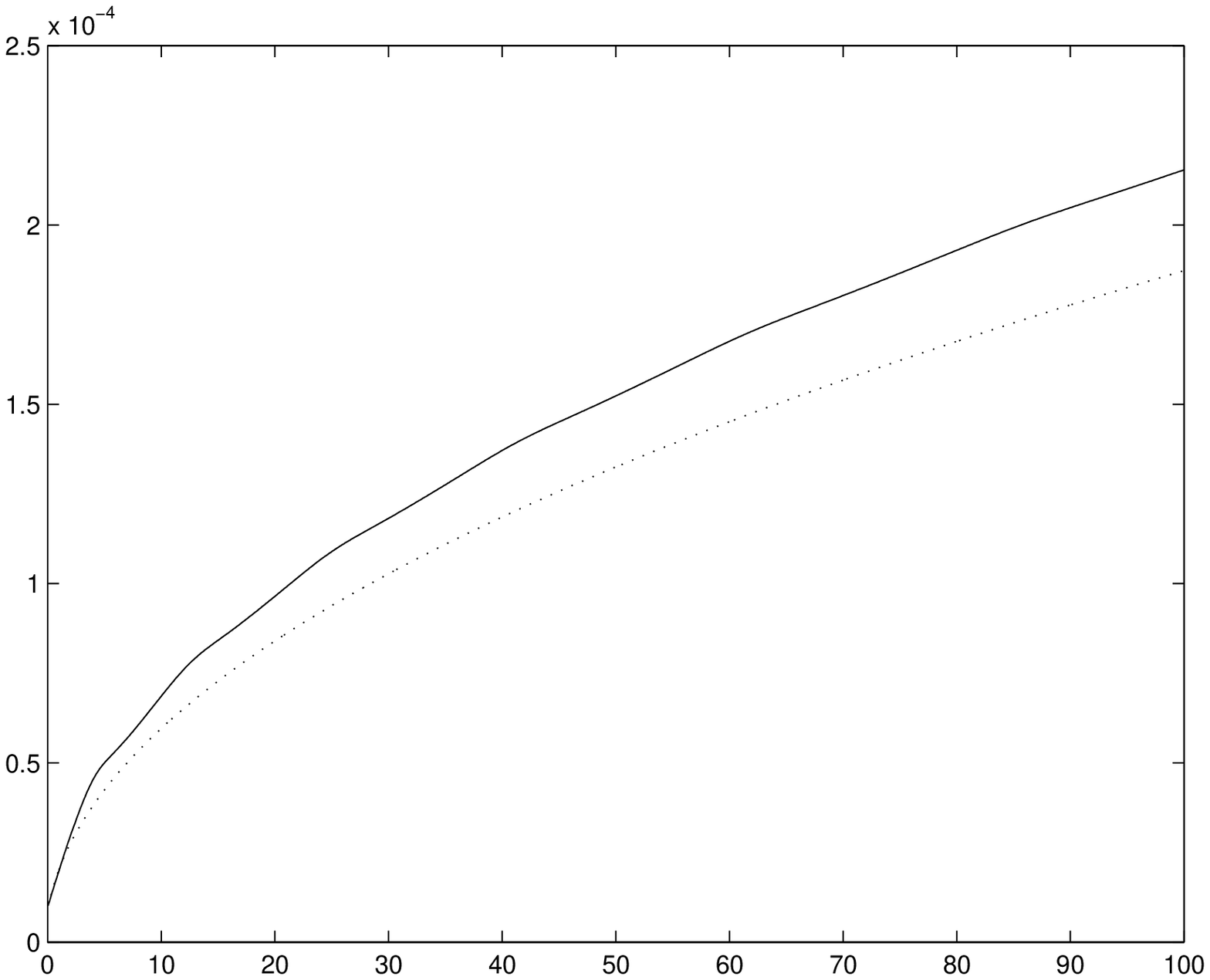}
\caption{The magnetic Weyl mode ${\cal H}_k(\tau)$, as for Fig.
5.}
\end{figure}
\begin{figure}
\includegraphics[height=3.2in,width=3.2in]{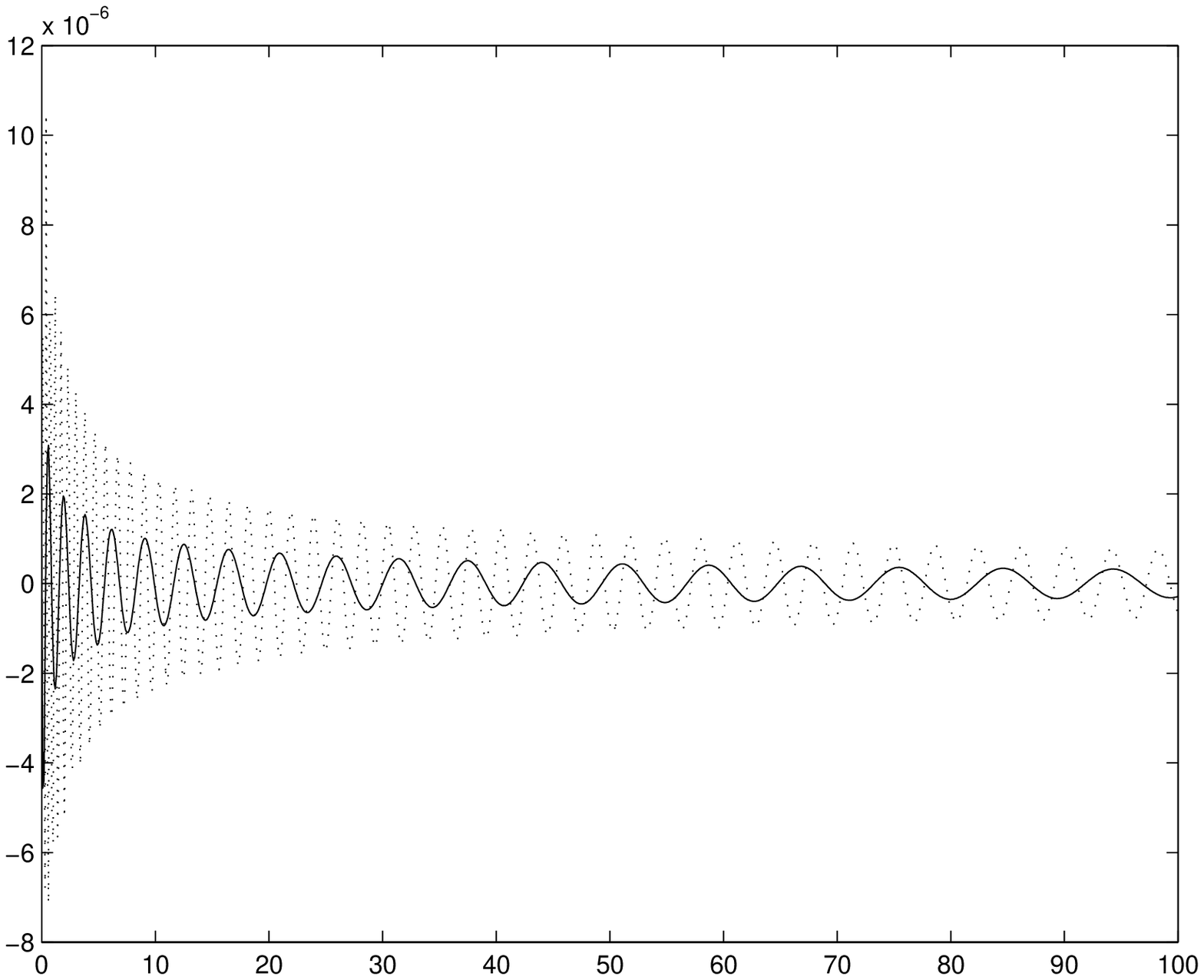}
\caption{The electric Weyl mode ${\cal E}_k(\tau)$ on sub-Hubble
scales, $k/(aH)_{in}=5$. The solid line is for the inflaton
potential $V={1\over4}\lambda^4\varphi^4$, while the dotted line
is for the corresponding time-averaged model, which behaves like
radiation.}
\end{figure}
\begin{figure}
\includegraphics[height=3.2in,width=3.2in]{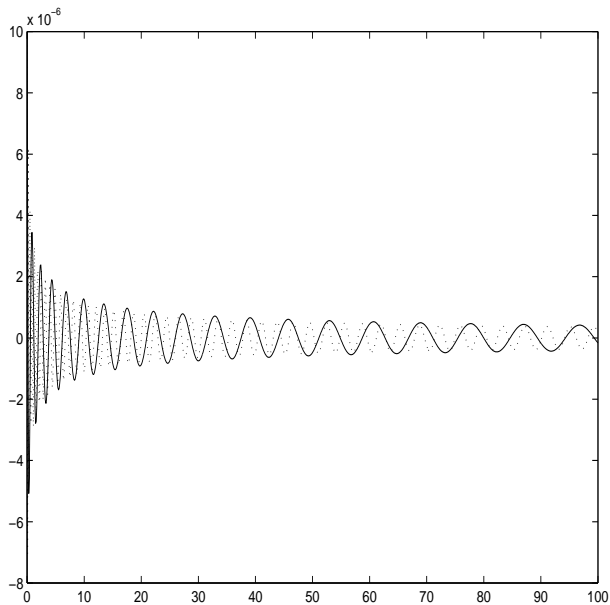}
 \caption{The magnetic Weyl mode ${\cal
H}_k(\tau)$, as for Fig. 7.}
\end{figure}

\section*{5. Conclusions}

Apart from specific features peculiar to the potential, these
figures reflect two key properties: (1) Super-Hubble gravitational
waves are not significantly amplified by coherent inflaton
oscillations during preheating; (2) gravitational waves on all
scales carry some imprint of the coherent oscillatory dynamics of
the inflaton during preheating. The latter point is clearly
brought out by a comparison with the evolution in the absence of
oscillations, i.e. for the time-averaged scale factor.

In particular, one can see that the electric Weyl modes on
super-Hubble scales, which determine the effect of gravitational
waves on the CMB (Challinor 2000), inherit oscillations from the
inflaton. In the early part of preheating, there is also some
amplification on average of ${\cal E}_k$, so that if backreaction
takes effect early, then this does produce a small amplification
relative to the no-oscillation time-averaged model. However, one
cannot expect these preheating imprints on ${\cal E}_k$ to lead to
detectable differences in the CMB power spectrum. Firstly, the
oscillations will effectively be averaged out, and secondly, any
amplification is likely to be scale invariant for all measurable
anisotropies, since the Hubble scale at preheating corresponds to
about 1 metre today, so that all cosmologically significant scales
at preheating effectively have $k=0$, and behave like the
particular scale chosen in our numerical integrations (Bassett et
al. 2000, Jedamzik and Sigl 2000).

Gravitational waves on sub-Hubble scales oscillate even in the
case of time-averaged background, but Figs. 3, 4, 7 and 8 show
that the frequency and amplitude of oscillation are significantly
modulated by the inflaton oscillations. Thus preheating leaves an
imprint on these scales. In principle, this could be detected, but
in practice the signal will be far too weak, since there is no
preheating amplification on these scales.

The absence of amplification is fundamentally due to the expansion
of the universe, since there is strong amplification on a
Minkowski background of gravitational waves during preheating
(Bassett 1997, Tilley 2000).
\\

{\bf Acknowledgement:} We thank Bruce Bassett for valuable
discussions and comments.

%\newpage
%\section*{References}

%\newpage

\end{document}